# Origin of enhanced visible-light photocatalytic activity of transition metal (Fe, Cr and Co) doped CeO$_2$: Effect of 3$d$-orbital splitting

Ke Yang, Dong-Feng Li, Wei-Qing Huang[*], Liang Xu, Gui-Fang Huang[#], Shuangchun Wen

*Department of Applied Physics, School of Physics and Electronics, Hunan University, Changsha 410082, China*

**Abstract**: Enhanced visible light photocatalytic activity of transition metal-doped ceria (CeO$_2$) nanomaterials have experimentally been demonstrated, whereas there are very few reports mentioning the mechanism of this behavior. Here we use first-principles calculations to explore the origin of enhanced photocatalytic performance of CeO$_2$ doped with transition metal impurities (Fe, Cr and Co). When a transition metal atom substitutes a Ce atom into CeO$_2$, $t_{2g}$ and $e_g$ levels of 3$d$ orbits appear in the middle of band gap owing to the effect of cubic ligand field, and the former is higher than latter. Interestingly, $t_{2g}$ subset of Fe$_{Ce}$ (Co$_{Ce}$ and Cr$_{Ce}$)-V$_o$-CeO$_2$ is split into two parts: one merges into the conduction band, the other as well as $e_g$ will remain in the gap, because O vacancy defect adjacent to transition metal atom will break the symmetry of cubic ligand field. These $e_g$ and $t_{2g}$ levels in the band gap are beneficial for absorbing visible light and enhancing quantum efficiency because of forbidden transition, which is one key factor for enhanced visible light photocatalytic activity. The band gap narrowing also leads to a redshift of optical absorbance and high photoactivity. These findings can rationalize the available experimental results and provide some new insights for designing CeO$_2$-based photocatalysts with high photocatalytic performance.

[*]. Corresponding author. *E-mail address:* wqhuang@hnu.edu.cn

[#]. Corresponding author. *E-mail address:* gfhuang@hnu.edu.cn

# 1. Introduction

Metal oxide photocatalysts have attracted increasing attention due to their potential applications in the environmental protection and energy utilization[1-6], such as the degradation of organic compounds and water splitting for hydrogen production. As one of the most reactive rare earth oxides, ceria ($CeO_2$) plays a very important role owing to its high chemical stability and photocatalytic activity, low cost, non-biological toxicity and environment friendly feature[7,8]. However, pure $CeO_2$ is a wide band gap semiconductor with a large gap of 3.20 eV, which limits its photocatalytic activity only under the ultraviolet (UV) radiation[9], which is about 5% solar energy available[10]. In order to extend its absorption edge into the visible (vis-) light region and improve its photocatalytic efficiency, various strategies have been proposed to engineer the band structure of $CeO_2$[5].

An effective strategy to modify the electronic structure of metal oxides (especially with wide band-gap), by shifting the valence band (VB) and/or conduction band (CB) edges at proper positions[11], or by forming impurity bands in the gap is to dope them with other elements[12,13]. The effectiveness of this approach for $CeO_2$ has been demonstrated by intensive experiments. As $CeO_2$ is doped with nonmetal elements (such as N, F, C), its band gap become narrower and the separation efficiency of photogenerated electron–hole pairs is high[14-19], thus enhancing its photocatalytic activity, even under vis-light region. Similarly, the introduction of various metal atoms into the $CeO_2$ lattice can also modify its electronic structure and photocatalytic activity[7,15,20-28]. Among the various metal elements, transition metal ions of multiple valency (such as Fe, Co, Y, Cu, Ni) have experimentally been doped into the $CeO_2$ lattice, which extends the optical absorption edge of $CeO_2$ to lower energies and enhances the mobility of excitons and facilitates surface reactions, thereby increasing the photocatalytic activity[26,28,29]. For example, Fe-doped $CeO_2$ shows higher photocatalytic activity than that of pure ceria in the presence of $H_2O_2$ owing less O vacancy defects[30]. The improved electron–hole separation and charge transfer process mechanism have been revealed via kinetics studies of the photocatalytic degradation of methyl orange over Fe–doped $CeO_2$ films based on the obtained XPS and UV–vis DRS results[24]. Co-doped $CeO_2$ nanorods exhibit high photocatalytic activity due to increasing O vacancy concentration and specific surface area. Moreover, it has been pointed out that Co ions act as photogenerated holes and electrons traps, which hinders the recombination of hole-electron pairs[23].

Oxygen vacancy defects, which are usually presented in impurity and dopant-controlled regimes of slightly substoichiometric pure or transition metals doped oxides, such as $TiO_2$ and $WO_3$, are of great importance as a separation or recombination center for the photogeneration electron–hole pairs[31-35]. In particular, a large of O vacancy defects have been found in a broom-like porous $CeO_2$ hierarchical structure due to the introduction of transition metal impurities, and demonstrated to be closely related to the enhancement of catalytic activity[36]. Similarly, Fe-doping can increase the concentration of O vacancies and specific surface of $CeO_2$, leading to an enhanced photocatalytic activity[30]. On the contrary, Yue *et al.* reported that the photocatalytic activity of transition metals (Ti, Mn, Fe and Co) doped $CeO_2$ reduces with decreasing O vacancy concentration[20]. Since the potential application prospect of doped $CeO_2$ and its composites for photocatalysis is highly inspirational, the clarification of its mechanism is therefore eagerly awaited due to these inconsistent experimental results. What is more, it has been demonstrated that the transition metal dopants and O vacancy greatly influence the photocatalytic performance of $CeO_2$, however, the effect of interaction between transition metal dopant and O vacancy defects on enhanced visible-light photocatalytic activity of $CeO_2$ is still not elucidated so far.

In this work, we systematically explore the physical mechanism on the enhanced photocatalytic activity of $CeO_2$ doped with transition metal elements, especially the effect of the interaction between transition metal dopant and O vacancy defects, by large-scale *ab initio* calculations. The various properties, such as relative stability, redox and mechanical properties, of pure $CeO_2$ have been intensively investigated by using first principles calculations[37-49], which are also widely used to study the properties of doped oxides and/or with O vacancy defects, such as $TiO_2$, $Cu_2O$, and $HfO_2$[11,13,50,51]. It is revealed that the reduced band gap and strong absorption induced by impurity levels are responsible for the enhanced visible-light photocatalytic activity of $CeO_2$ doped with nonmetal impurity[52]. To explore the effects of the transition metal dopants and O vacancy, Fe, as a dopant, has been chosen because of their ionic radii smaller than Ce, easily replacing Ce lattice, which have been experimentally found that Fe doping oxides photocatalyst, especially in TiO$_2$ [53] and CeO$_2$ [9,20,24,25], have enhanced photocatalytic activity. Cr and Co dopants, near to Fe in period table, have also been considered in order to reveal the origin of photocatalytic performance of transition metal doped CeO$_2$. The calculated results show that enhanced visible-light photocatalytic activity of doped CeO$_2$ can mainly be attributed to 3$d$-orbital splitting of transition metal atom and the reduced band gap. The interaction between transition metal dopant and O vacancy defect influences the positions of $t_{2g}$ and $e_g$ levels of 3$d$ orbits because the symmetry of cubic ligand field is broken. These results offer a physical interpretation for the available experimental results and are useful for designing and understanding CeO$_2$-based photocatalysts.

## 2. Theoretical model and computational details

Bulk cubic CeO$_2$ is fluorite-type (face-centered cubic, Fm3-m, Ce-O bond length 2.34 Å), and Ce and O atoms are eight- and four-fold coordinated, respectively. All of calculations have been conducted in a 2 × 2 × 2 supercell (including 96-atom) built from the conventional 12-atom cubic unit cell of CeO$_2$ (see Fig. S1), in order to reduce the dopant–dopant and vacancy–vacancy interactions. Six models have been discussed: Fe$_{Ce}$-CeO$_2$, Cr$_{Ce}$-CeO$_2$, and Co$_{Ce}$-CeO$_2$ (one transition metal atom (Fe, Cr, Co) substituting a Ce atom), Fe$_{Ce}$-V$_o$-CeO$_2$, Cr$_{Ce}$-V$_o$-CeO$_2$, and Co$_{Ce}$-V$_o$-CeO$_2$ (an O vacancy is introduced in CeO$_2$ through removing one of the O atoms nearest neighbor impurity atom (Fe, Cr, and Co)). This is because O vacancy is preferred to be created at the nearest neighbor site of the dopant[49]. For comparing and analyzing the influence of interaction between O vacancy and dopants, V$_o$-CeO$_2$ (only O vacancy defect in CeO$_2$) are also considered.

In this work, all of calculations have been performed using the first-principles density functional theory (DFT) within the projector augmented wave (PAW) methods[54], as implemented in the Vienna *ab initio* Simulation Package (VASP) [55,56]. Various density-functional theory methods[38,57, 58], including the local spin density approximation (LSDA), local density approximation (LDA), LDA+U, generalized gradient approximations (GGA), Perdew−Burke−Ernzerhof (PBE0), and Heyd−Scuseria−Ernzerhof (HSE) hybrid functional have been tested to investigate the properties of cubic CeO$_2$. Among them, LDA and GGA fail to reproduce correct electronic structure, due to an overestimation of the 4f-electron delocalization related to the self-interaction problem. Hybrid approaches, including a fraction of nonlocal Hartree-Fock (HF) exchange in the exchange-correlation functional, are not parameterized on a specific property and are thus in principle more generally applicable, especially as concerns the demanding $CeO_2$ case. However, these accurate methods are high costly and time consuming, extremely restricting their real applications. Fortunately, the simple, empirical and weakly-demanding computational approach to obtain appropriate band structure is to add an on-site Coulombic repulsion term, leading to the so-called DFT+U approach, which has shown a wide use in studying bulk $CeO_2$ and its composite[58,59]. For $CeO_2$, the calculated structural parameters, band gap, and the levels in the vicinity to Fermi level obtained from LDA+U, PBE0, and HSE06 are in agreement with each other and coincided with experimental results[38,58,59].

Considering more than one system discussed here, we choose the DFT/LDA+U method. We have performed extensive tests to determine the appropriate Hubbard U parameters (Ce 4f and O 2p are 9.0 and 4.5 eV, respectively), which reproduced the correct lattice parameters (5.40 Å) and energy gap (3.2 eV) for cubic $CeO_2$. Moreover, the position and width of the Ce 4f and O 2p bands are in good agreement with the experiment and previous calculations[52,59]. This indicates that the DFT/LDA+U computations can correctly characterize the Ce 4f and O 2p states. The valence atomic configurations are Ce: $4f^1 5d^1 6s^2$, O: $2s^2 2p^4$, Fe: $3d^7 4s^1$, Cr: $3d^5 4s^1$, and Co: $3d^8 4s^1$, respectively. The plane wave cut off is set to 500 eV, and the k mesh of 3×3×3 and 5×5×5 for the calculated systems are used for geometry optimizations and calculating the density of states using the Monkhorst-Pack scheme, respectively. In the geometrical optimization, total energy and all forces on atoms are converged to less than $10^{-6}$ eV and 0.03 eV/Å. All these parameters are enough to guarantee accurate results.

## 3. Results and discussion

## 3.1 Geometric Structures

To assess the accuracy of our computation method, we have performed a series of calculations for pure CeO$_2$. First, we perform structural optimization for pristine CeO$_2$. The equilibrium lattice constant after optimization is *a*=*b*=c=5.406 Å, which is in good agreement with the experimental value (*a*=*b*=c=5.407 or 5.411 Å) [60], assuring the validity of the calculations. Consequently, the same calculation conditions are adopted to calculate the geometrical and electronic structures of the pure and doped CeO$_2$.

The changes in lattice parameters and distance between adjacent atoms of doped CeO$_2$ systems are summarized in Table I. Compared with pure CeO$_2$, the optimized lattice constants of transition doped CeO$_2$ decrease from 10.812 Å to 10.680 Å, 10.681 Å, and 10.683 Å for Fe$_{Ce}$-CeO$_2$, Cr$_{Ce}$-CeO$_2$, and Co$_{Ce}$-CeO$_2$, respectively. The reduction of lattice constants is attributed to fact that the distance of Fe–O (2.116 Å, Cr–O: 2.124 Å, Co–O: 2.137 Å) is shorter than Ce-O bond (2.343 Å), because the electronegativity of Fe atom is bigger (1.83, Cr: 1.66, Co: 1.88) compared with that of Ce atom (1.12) [61], and the ionic radii of Fe$^{3+}$ (0.55 Å), Cr$^{3+}$ (0.62 Å) and Co$^{3+}$ (0.54 Å) is smaller than the Ce$^{4+}$ radius (0.90 Å) [62]. However, the equilibrium lattice constant of Fe$_{Ce}$-V$_o$-CeO$_2$ (10.775 Å, Cr$_{Ce}$-V$_o$-CeO$_2$: 10.809 Å, Co$_{Ce}$-V$_o$-CeO$_2$: 10.785Å) is larger than that without O vacancy. In order to understand the lattice variation, the lattice pure CeO$_2$ with an O vacancy (10.821 Å) have been discussed, which is larger than pure CeO$_2$ (10.812 Å) because Ce atoms adjacent to O vacancy are pulled away from O vacancy site. Therefore, compared with Fe$_{Ce}$ (Cr$_{Ce}$ and Co$_{Ce}$)-V$_o$-CeO$_2$, the lattice constant of Fe$_{Ce}$ (Cr$_{Ce}$, Co$_{Ce}$)-V$_o$-CeO$_2$ decreases as a result of the appearance of O vacancy. With the effect of interaction between O vacancy and transition metal dopant, the distances between metal ionic and surrounding O atoms are unequal and the local symmetry are broken.

## 3.2. Defect formation energy

The analysis on optimized geometrical structure shows that the different dopants generate different lattice distortions, which makes the supercells of doped CeO$_2$ with various total energies. In order to examine the relative stability of different doped systems, we have calculated the formation energies ($\Delta E_{form}$) of the O vacancy, doped CeO$_2$ with and without O vacancy according to the following equation, respectively.

The formation energy $E_{vac}$ of the O vacancy:

$$E_{vac} = E_{tot}(vac) - E_{tot}(CeO_2) + \sum n_O \mu_O \quad (1)$$

The formation energy $\Delta E_{form}$ of the doped $CeO_2$ without O vacancy:

$$\Delta E_{form} = E_{tot}(\alpha) - E_{tot}(CeO_2) - \sum n_X \mu_X + \sum n_{Ce} \mu_{Ce} \quad (2)$$

The formation energy $\Delta E_{form-vac}$ of the doped CeO₂ with O vacancy:

$$\Delta E_{form-vac} = E_{tot}(\alpha - vac) - E_{tot}(vac) - \sum n_X \mu_X + \sum n_{Ce} \mu_{Ce} \quad (3)$$

where X=Fe, Cr, or Co, $E_{tot}(vac)$, $E_{tot}(\alpha)$ and $E_{tot}(\alpha - vac)$ is the total energy of $V_o$-$CeO_2$, $Fe_{Ce}$ ($Cr_{Ce}$ and $Co_{Ce}$)-$CeO_2$ and $Fe_{Ce}$($Cr_{Ce}$ and $Co_{Ce}$)-$V_o$-$CeO_2$, respectively. $E_{tot}(CeO_2)$ is the total energy of the corresponding pure $CeO_2$, while $\mu_O$, $\mu_X$ and $\mu_{Ce}$ are the chemical potentials of removed atom O, the substitutional atom X and the substituted Ce host atom, respectively. The chemical potentials of the X dopants are calculated with respect to bulk ($X_2$ dimers) reference levels for Ce substitution. The $\mu_{Ce}$ ($\mu_O$) is the total energy per atom of Ce (O) in its reference phase. For Ce the reference phase used is the bulk metallic face-centered-cubic (fcc) structure, whereas for O it is the $O_2$ molecule (we note that the choice of reference phase is not unique). The value of $E_{form}(CeO_2)$ is then calculated to −11.94 eV in DFT/LDA+U, which coincides with −11.33 eV[49,58]. Table I lists the formation energies of all the doped CeO₂ systems studied. As one metal element (Fe, Cr, or Co) substituting a Ce atom, it produces one or more than one hole below the Fermi Level, thus causing it to act as an acceptor. The positive formation energy indicates that it needs energy (3.57 eV) to create an O vacancy in CeO₂ lattice[44,49]. The formation energy is 2.46, -1.09 and 4.39 eV for Fe, Cr and Co doped into CeO₂, respectively. As is known, the smaller defect formation energy is more preferable to incorporate impurity atoms into the host material. Among these transition metal elements, Cr atom is most likely to be doped into. When O vacancy is introduced, the Fe (Cr, Co) is easier to be doped due to small formation energy than that without O vacancy, agreeing with those experimental results[9,21,23-25,63]. This is similar with the cases for other metal elements (such as Pt, Rh) doped CeO₂[45].

## 3.3 Electronic properties

Before discussing the influence of dopants on the electronic properties of matrix, it is necessary to briefly review the crystal field theory in order to understand the role of dopants in $CeO_2$ well. Here, we mainly focus on the 3$d$ orbital splitting in a cubic ligand field. In the ligand field picture of a complex system, the ligands surrounding the central ion are regarded simply as providing an electrostatic field with which the central ion electrons can interact. The electron distribution of each ligand atom closest to the central ion is replaced by a point charge or dipole at an effective distance from the outer electrons of the ion. It is well-known that, in a spherically symmetric field of negative charges, $d$ orbits of metal remain degenerate, but all of them are raised in energy as a result of the repulsion between the negative charges on the ligands and the central metal atom. However, rather than in a spherical field, the ligands in a cubic field are allowed to interact with the metal atom. Thus, the degeneracy of the $d$ orbitals is removed, splitting into two subsets. One is a subset of two orbitals (namely $d_{z2}$ and $d_{x2-y2}$), which are basis to the irreducible representation $e_g$, another is a subset of three orbitals (namely $d_{xy}$, $d_{yz}$ and $d_{zx}$), which are basis to the irreducible representation $t_{2g}$, while $e_g$ is lower than $t_{2g}$ (see Fig. S2 (a)) [64,65]. To understand the mechanism of the splitting of $d$ orbitals in a cubic ligand field, it is necessary to present pictures showing the shapes of the usual five $d$ orbits, Fig. S2 (b). Imagine eight ligands lying at corners of a cube to form a cubic geometry. As the metal ion is at the center of the cube, its $d_{xy}$, $d_{xz}$, and $d_{yz}$ orbitals point toward the ligands, resulting into strong interaction between these orbits and ligands; whereas its $d_{x2-y2}$ and $d_{z2}$ orbits lie between the ligands, thus the interaction between these orbits and ligands is relative weak. As a consequence, the energy of the $d_{xy}$ ($d_{xz}$ and $d_{yz}$) orbit is higher than the $d_{x2-y2}$ and $d_{z2}$ orbitals, and the splitting phenomenon ($t_{2g}$ and $e_g$) can be observed in a cubic field.

Band structure, density of states (DOS) and partial density of states (PDOS) of pure $CeO_2$ are shown in Fig. S3 (a1) and (a2). The position and width of the Ce 4$f$ and O 2$p$ bands are in good agreement with the experiment and previous calculations[42,58,59]. Here, the calculated band structure, electron density distributions, DOS and PDOS of $Fe_{Ce}$-$CeO_2$ are displayed in Fig. 1 (a), ($sp$), ($e_g$) and ($t_{2g}$), Fig. 2 (a1-a3) and Fig. 3 (a), respectively. The Fermi level is set at zero energy to easily identify the band gap and relative position of the states from the transition metal impurity atom. Fig. 1 (a) shows that Fe doping leads to a narrow of band gap (~2.75 eV) of $CeO_2$, which is agreement with experimental value 2.81 eV[20]. Both the valence band (VB) and conduction band (CB) shift negatively, relative to Fermi level, and the movement of latter is 0.45 eV larger than that of former. Compared to pure $CeO_2$, the decrease of band gap is mainly owing to the moving down of CB, which is one of the main reasons for enhanced photocatalytic activity of Fe-doped $CeO_2$. Fig. 3 (a) shows that one $4s$ and three $4p$ orbits have contributions to the bottom of CB, demonstrating a $sp3$ hybridization feature. Compared with $3d$ electron of transition metal, the $4s4p$ orbitals of Fe atom are more inclined to form bonding with $2p$ orbits of eight O atoms surrounding Fe atom, which can be clearly seen from the electron density distributions of $4s4p$ orbits, as shown in Fig. 1 ($sp$). These $sp3$ hybrid orbits are bonding states with lower energies, thus merging into the VB (see Figs. 1 (a) and (b)). We can consider that Fe atom is located in a cubic symmetric field of negative charges, and the eight O atoms around Fe act as the ligands.

In the cubic ligand field, $3d$ orbits of Fe atom have split into higher $t_{2g}$ levels lying at 0.74 eV (including three levels) and lower $e_g$ levels lying at near Fermi level (including two levels) in the gap (see Figs. 1 (a) and (b)). The lower $e_g$ levels near Fermi level are occupied by four valence electrons, while the higher $t_{2g}$ levels are unoccupied. Fig. 2 (a3) and Fig. 3 (a) display that the $e_g$ subset made of Fe $3d_{z2}$, $3d_{x2-y2}$ states and few O $2p$ states are basically non-bonding due to local $3d$ electrons, acting as electron donors. The hybridization of O $2p$ states is due to the fact that the O atoms obtain less electrons from Fe $4s4p$ orbits in doped $CeO_2$ compared to Ce in pure $CeO_2$ because the electronegativity of Fe is larger than that of Ce. For the unoccupied $t_{2g}$ (including $d_{xy}$, $d_{yz}$ and $d_{xz}$) subset, a few O $2p$ and Ce $4f$ $5d$ orbits form anti-bonding states, acting as electron acceptors (see Fig. 1 ($t_{2g}$) and Fig. 2 (a2) (a3) and Fig. 3 (a)).

It has been demonstrated that the photocatalytic activity of a photocatalyst is not only dependent on the quantity of available photons, but also on the quantum efficiency. During the photocatalytic reaction process, the recombination of photogenerated carriers easily occurs, thus reducing the quantum efficiency. Considering the appearance of $e_g$ and $t_{2g}$, the Fe atom can act as photogenerated electron and hole separatation or recombination centers. However, it is quite difficult that the electron in $e_g$ levels leap into $t_{2g}$ levels, or vice versa, because of forbidden transition. Firstly, the electrons cannot leap from one Fe atom to another when the distance between two nearest

Fe atoms is long enough (such as about 10.68 Å). Secondly, the $e_g$ and $t_{2g}$ subsets are mainly constituted of $d_z^2$, $d_{x2-y2}$ and $d_{xy}$, $d_{xz}$, $d_{yz}$ orbits, their main momentum quantum number ($n$) and angular momentum quantum number ($l$) are the same for a Fe atom. Since the difference of angular momentum quantum number on two states $\Delta l=\pm 1$ is one of the nessarry conditions when the electrons leap from a state to another state, there are only three pathways that the electrons in VB transfer to CB in Fe-doped CeO$_2$ (see sketches of the three black arrows in Fig. 1(a)). The first pathway is that an electron leaps from VB to CB by capturing a photon with energy least $E_1$= 2.75 eV. The second pathway is that an electron from VB firstly moves to $t_{2g}$ levels by trapping a photon with $E_2$=1.96 eV, then it leap to CB. The last one is that an electron from VB firstly moves to $e_g$ levels, then it leap to CB by trapping a photon with $E_3$=1.76 eV. These transitions indicate that during the photocatalytic reaction, the electron in the impurity levels ($e_g$ and $t_{2g}$ levels) moves to CB by capturing a photon with enough energy, meanwhile a hole is left in the $e_g$ and $t_{2g}$ levels. Then the electron from VB can jump to the $e_g$ and $t_{2g}$ levels by capturing another appropriate photon, which means that the hole in the $e_g$ and $t_{2g}$ levels is eliminated and new hole forms in the top of VB. As a result, the $e_g$ and $t_{2g}$ levels do not change, while an electron and a hole are generated in CB and VB, respectively, indicating the separation of photogenerated electron-hole pairs. Therefore, the $e_g$ and $t_{2g}$ levels in Fe-doped CeO$_2$ are beneficial to enhancing quantum efficiency, which is one key factor for its high photocatalytic activity, even under vis-light region.

The 3$d$-oribtal splitting of Fe atoms in the cubic ligand field is expected to be general in transition metal doped CeO$_2$. The Cr and Co atoms, as two paradigms, are chosen to study the interaction of dopants and host atoms. The electronic properties of Cr and Co-doped CeO$_2$ have been also calculated using DFT, which show similar results (Figs. 2-4). The band gaps of Cr and Co-doped CeO$_2$ decrease to 2.77 eV and 2.81 eV, respectively, which is agreement experimental value (2.84 eV for Co-doped CeO$_2$ [63]). From the Fig. 4, it can be clearly seen that the $t_{2g}$ and $e_g$ levels of Cr$_{Ce}$-CeO$_2$ is higher than that of Co$_{Ce}$-CeO$_2$. This is due to the fact that there are only three electrons occupy in $e_g$ levels of Cr$_{Ce}$-CeO$_2$, while four electrons lie at $e_g$ levels and the other electron takes up one of $t_{2g}$ levels of Co$_{Ce}$-CeO$_2$. The $e_g$ and $t_{2g}$ levels of Cr$_{Ce}$-CeO$_2$ and Co$_{Ce}$-CeO$_2$ are mainly composed of 3$d_z^2$, 3$d_{x2-y2}$, and 3$d_{xy}$, 3$d_{xz}$, 3$d_{yz}$ states, maxing with small O 2$p$, Ce 4$f$ and 5$d$ states (see Figs. 2-4), similar with the case of $Fe_{Ce}$-$CeO_2$. The $t_{2g}$ subset of $Cr_{Ce}$-$CeO_2$ and $e_g$ subset of $Co_{Ce}$-$CeO_2$ are very near to top of CB and bottom of VB, respectively, which will act as acceptor and donor, respectively. The electronic transition is also forbidden between $e_g$ and $t_{2g}$ levels for $Cr_{Ce}$-$CeO_2$ and $Co_{Ce}$-$CeO_2$, and the energy gaps between different subsets have also been given in the table II. These impurity levels are responsible for the red-shifted optical absorption and enhanced photocatalytic activity of doped $CeO_2$.

The transition metal doping can increase the concentration of O vacancy in $CeO_2$ has been experimentally demonstrated[30,36], which is further verified by the formation energy calculations in Section 3. 2. In order to reveal the effect of interaction between metal impurity and O vacancy on enhanced visible-light photocatalytic activity of $CeO_2$, the electronic properties of $V_o$-$CeO_2$ with and without metal dopants have also been calculated, as shown in Fig. S3 (b1 and (b2), and Figs. 5 -8. The presence of O vacancy reduces the band gap of $CeO_2$, due to two Ce electrons lying at the impurity near CB. The levels originate from hybridized Ce $5d$ and small $4f$ states. The charge density distributions (Fig. S4) show two Ce electrons are located at O vacancy site which is agreement with previous reports[47]. Previous investigations[45,61] have suggested that it is most possible that metal impurity replaces Ce nearest O vacancy site, because the metal atomic valence electrons are perturbed in unsymmetrical static electric field, as shown in Fig. 5. Various Fe $3d$ orbits can be clearly seen from Fig. 5 when Fe dopant and O vacancy coexist in $CeO_2$. The $t_{2g}$ components have split into two subsets: one subset includes one level lying at middle of band gap, another subset are two levels lying at CB (see Fig. 5 (b)), while the $e_g$ subset is still located in the middle of band gap (see Fig.5 and Fig. 7 (a)). Unexpectedly, there are six electrons occupy the $e_g$ levels and higher level $t_{2g}$ (Fig. 5(b)), which are two more than those in the case of $Fe_{Ce}$-$CeO_2$. This may be attributed to two factors: 1) partial $4s4p$ levels shift up to CB due to disturbing the bonding states between Fe $4s4p$ and O $2p$ when O vacancy is introduced (see Fig.5 (sp) and Fig.7 (a)), 2) partial $t_{2g}$ levels shift down, because the coulomb repulsion of $t_{2g}$ components in $[\bar{1}\ 1\ \bar{1}]$ direction of ligands becomes weaken in broken symmetrical cubic ligand field.

Bottom panels of Figs. 5 display the electron and hole density distributions with an isovalue of 0.008 e/Å3, in which ($sp$), ($e_g$) and ($t_{2g}$) is $4s4p$, $e_g$ and $t_{2g}$ impurity levels of $Fe_{Ce}$-$V_O$-$CeO_2$, respectively. It is obvious that the interactions between Fe and near O are not symmetrical. Surprisingly, the lower energy $t_{2g}$ level is mainly composed of Fe 3$d$ electrons hybridized with some Ce 4$f$ and 5$d$ states, and the electrons located at O vacancy site have disappeared (Figs. 5 ($t_{2g}$) and 7 (a)). Using the Bader atomic population analysis, it is found that the Fe atom in Fe$_{Ce}$-V$_o$-CeO$_2$ loses 0.214 e less than that in Fe$_{Ce}$-CeO$_2$, which is the result of charge compensation between Fe and O vacancy. This is one reason for the lower formation energy of Fe$_{Ce}$-V$_o$-CeO$_2$ than Fe$_{Ce}$-CeO$_2$ (Table I). For the Cr$_{Ce}$-V$_o$-CeO$_2$ and Co$_{Ce}$-V$_o$-CeO$_2$, the calculations show similar results (Fig. 6, 7 and 8). Compared to pure CeO$_2$, the band gaps of Cr$_{Ce}$-V$_o$-CeO$_2$ and Co$_{Ce}$-V$_o$-CeO$_2$ are decreased to 2.71 eV and 2.70 eV, respectively. Four and seven electrons occupy in $e_g$ and $t_{2g}$ for Cr$_{Ce}$-V$_o$-CeO$_2$ and Co$_{Ce}$-V$_o$-CeO$_2$, respectively. For the Co$_{Ce}$-V$_o$-CeO$_2$, all of $t_{2g}$ components present in the band gap due to lower Co 3$d$ orbitals energy compared with Fe and Cr (see Figs. 6 and 7). Considering forbidden transition, different energy gaps between different subsets have listed in Table II. Based on the above analysis, it is expected that transition metal doped CeO$_2$ with less O vacancies is more favorable to absorb vis-light because of wider absorption range and less electrons in delocalized 3$d$ states.

## 3.4 Optical properties

Besides a low recombination rate of photogenerated carriers, the strong light absorption is another fundamental premise for a high-efficiency photocatalyst. As is well known, the dielectric function of the semiconductor materials is mainly connected with the electronic response. The imaginary part $\varepsilon_2$ of the dielectric function $\varepsilon$ is calculated from the momentum matrix elements between the occupied and unoccupied wave functions, given by:

$$\varepsilon_2 = \frac{ve^2}{2\pi\hbar m^2 \omega^2} \int d^3k \sum_{n,n'} |\langle kn|p|kn'\rangle|^2 f(kn)(1-f(kn'))\delta(E_{kn}-E_{kn'}-\hbar\omega) \quad (4)$$

where $\hbar\omega$ is the energy of the incident photon, p is the momentum operator $(\hbar/i)(\partial/\partial x)$, ($|kn\rangle$) is a crystal wave function and f(kn) is Fermi function. The real part $\varepsilon_1$ of the dielectric function $\varepsilon$ is evaluated from the imaginary part $\varepsilon_2$ by Kramer–Kronig transformation. The absorption coefficient I($\omega$) can be derived from $\varepsilon_1$ and $\varepsilon_2$, as following:

$$I(\omega) = \sqrt{2}\omega \left[\sqrt{\varepsilon_1^2(\omega) + \varepsilon_2^2(\omega)} - \varepsilon_1(\omega)\right]^{1/2} \quad (5)$$

which depends on $\varepsilon_1$ and $\varepsilon_2$ and thus on the energy. All other optical constants can also be obtained. The relations above are the theoretical basis of band structure and optical properties analyzing which reflected the mechanism of absorption spectral caused by electronic transition between different energy levels.

The absorption spectra of pure and doped $CeO_2$ are calculated by the Fermi golden rule within the dipole approximation, and presented in Fig. 9. The pure $CeO_2$ shows an onset of optical absorption about 390 nm, which is attributed to the intrinsic transition from O $2p$ to CB (~3.2 eV). The absorption edge of the $Fe_{Ce}$-$V_O$-$CeO_2$, $Cr_{Ce}$-$V_O$-$CeO_2$ and $Co_{Ce}$-$V_O$-$CeO_2$ has a small red shift (50~70 nm) compared to that of the pure $CeO_2$, which is attributed to the transition from VB to CB, in agreement with experimental results[28,63]. However, the $Fe_{Ce}$-$CeO_2$, $Cr_{Ce}$-$CeO_2$ and $Co_{Ce}$-$CeO_2$ have strong absorption in the entire vis-region coincided with experimental results[20], especially resonant-like absorption peak near 600 nm, indicating that these nanomaterials are promising efficient vis-active photocatalysts. This can be attributed to the appearance of $3d$ intermediated bands in the band gap (Figs.1-4). Remarkably, transition metal impurity (Fe, Cr, Co) doped $CeO_2$ without O vacancy have a stronger absorption compared to that with O vacancy. Especially, the $Fe_{Ce}$-$CeO_2$ has a big absorption peak near 600 nm, which its absorption is about twice and four times than $Co_{Ce}$-$CeO_2$ and $Cr_{Ce}$-$CeO_2$. Thus, it should be suggested strongly to synthesize transition metal-doped $CeO_2$ with less O vacancy by controlling experimental condition, such as increasing of oxygen flow[20].

## 4. Summary

In summary, a systemic first-principles calculation has been conducted to study the effect of transition metal doping on the structural, electronic and optical properties of doped $CeO_2$ with and without O vacancy. The introduction of Fe (Cr, Co) will reduce the band gap of $CeO_2$, about 2.7~2.8 eV. Moreover, the transition metals introduce $3d$ states in the middle of band gap of $CeO_2$. The $3d$ states act an important role on photon-generated carrier separation. Without O vacancy, the $3d$ orbits splitting of transition metal impurity in symmetrical cubic ligand field is more benefit for photocatalytic performance in $CeO_2$. The O vacancy do not seem conducive to improve the photocatalytic activity of $CeO_2$. This work reveals the mechanism on the enhanced vis-light photocatalytic activity of transition metal-doped $CeO_2$ and enrich our understanding of the $CeO_2$-based composites for developing photocatalysts with high activity.

**Supplemental Materials**

See supplemental material at http://link.springer.com/journal/339 for geometry structure of doped $CeO_2$ Fig. S1, electronic properties of pure $CeO_2$ Figs. S2 and S3, and schematic plots of electron density distribution of the 3d orbits Fig. S4, respectively.

**Acknowledgements**

This work is partially supported by the National Natural Science Foundation of China (Grant Nos. 11574079) and the Undergraduate Students Research Program of School of Physics and Electronic, Hunan University (No. USRP201609).

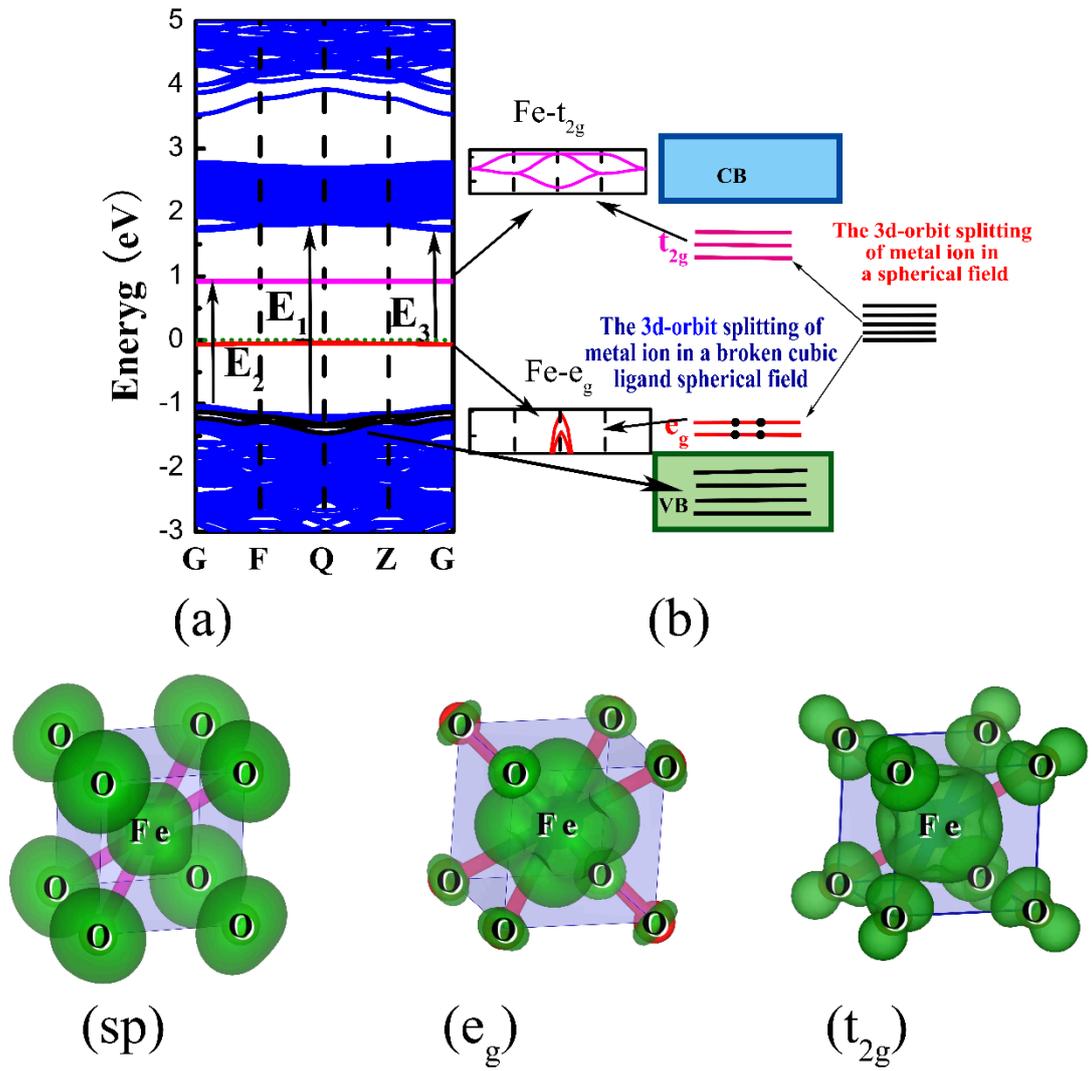

**Fig. 1** (a) Band structure of $Fe_{Ce}$-$CeO_2$. The Fermi level is set at 0 eV. The red and magenta lines denote the $e_g$ and $t_{2g}$ levels, respectively. (b) Schematic plots Fe-3$d$-orbits splitting for $Fe_{Ce}$-$CeO_2$. Bottom panels show the electron and hole density distributions with an isovalue of 0.008 e/Å3: ($sp$), ($e_g$) and ($t_{2g}$) are 4$s$4$p$, $e_g$ and $t_{2g}$ impurity levels of $Fe_{Ce}$-$CeO_2$, respectively.

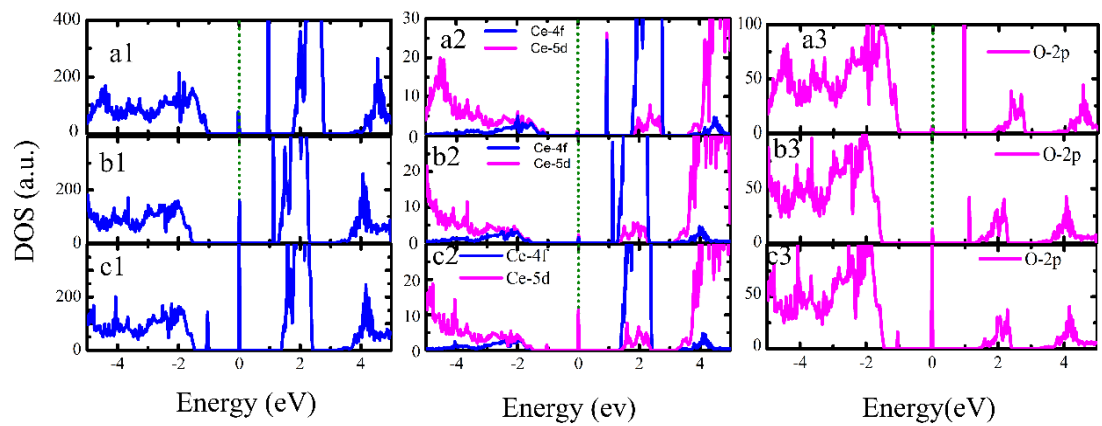

**Fig. 2** DOS and PDOS for Fe-CeO$_2$(a1-a3), Cr-CeO$_2$(b1-b3), and Co-CeO$_2$(c1-c3). The Fermi level is set at 0 eV.

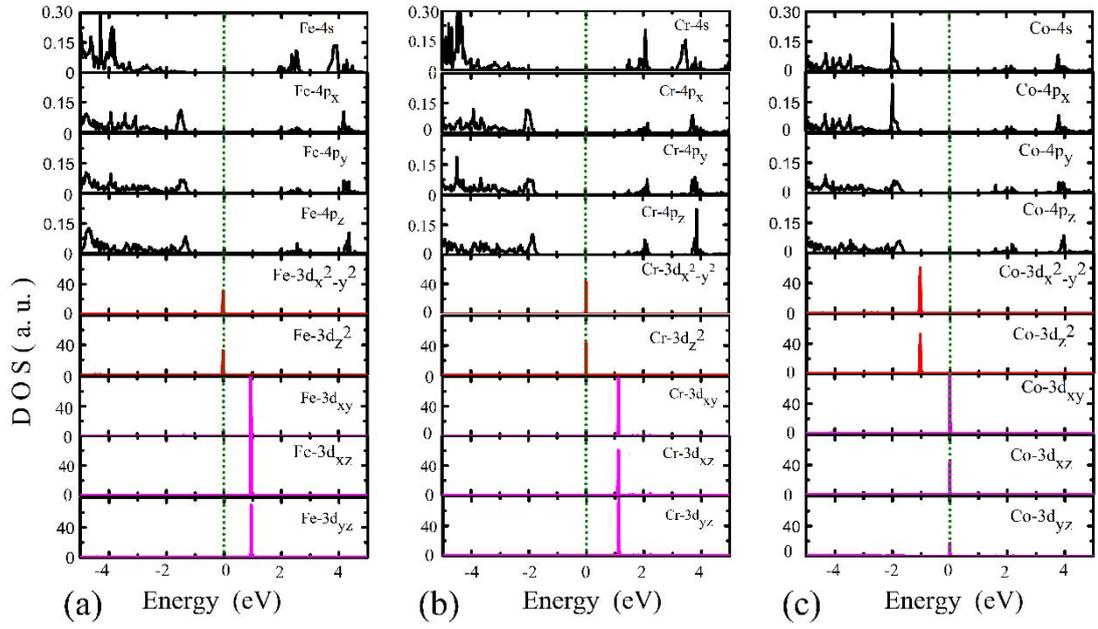

**Fig. 3** PDOS of impurity atoms in $Fe_{Ce}$-$CeO_2$ (a), $Cr_{Ce}$-$CeO_2$ (b), and $Co_{Ce}$-$CeO_2$ (c). The Fermi level is set at 0 eV.

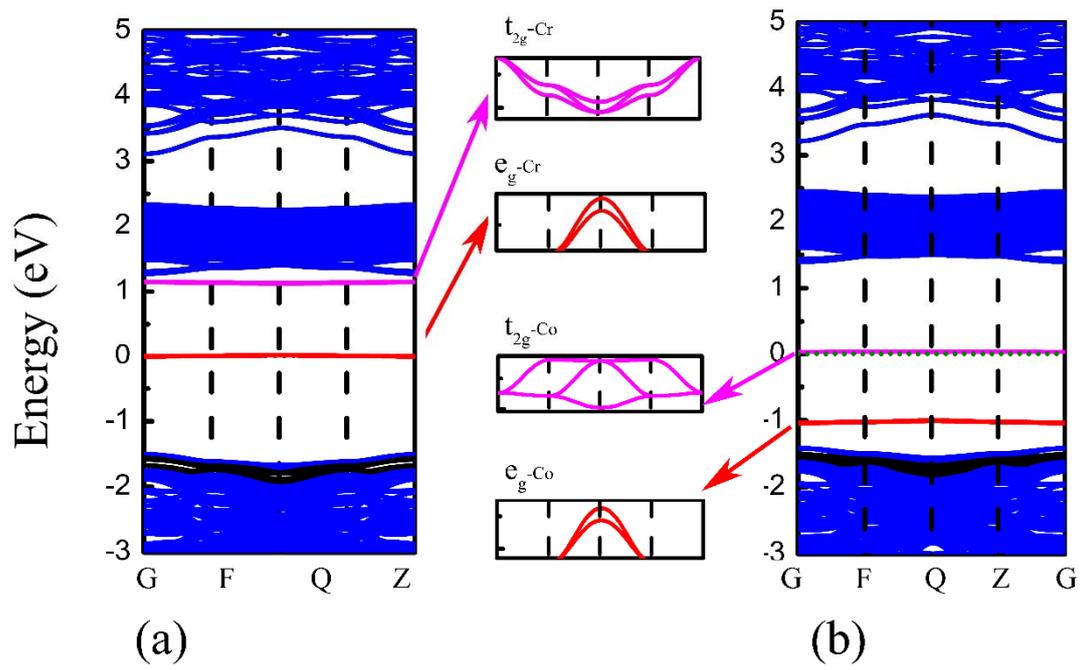

**Fig. 4** Band structures of (a) $Cr_{Ce}$-$CeO_2$ and (b) $Co_{Ce}$-$CeO_2$. The Fermi level is set at 0 eV. The red and magenta lines denote the $e_g$ and $t_{2g}$ impurity levels, respectively.

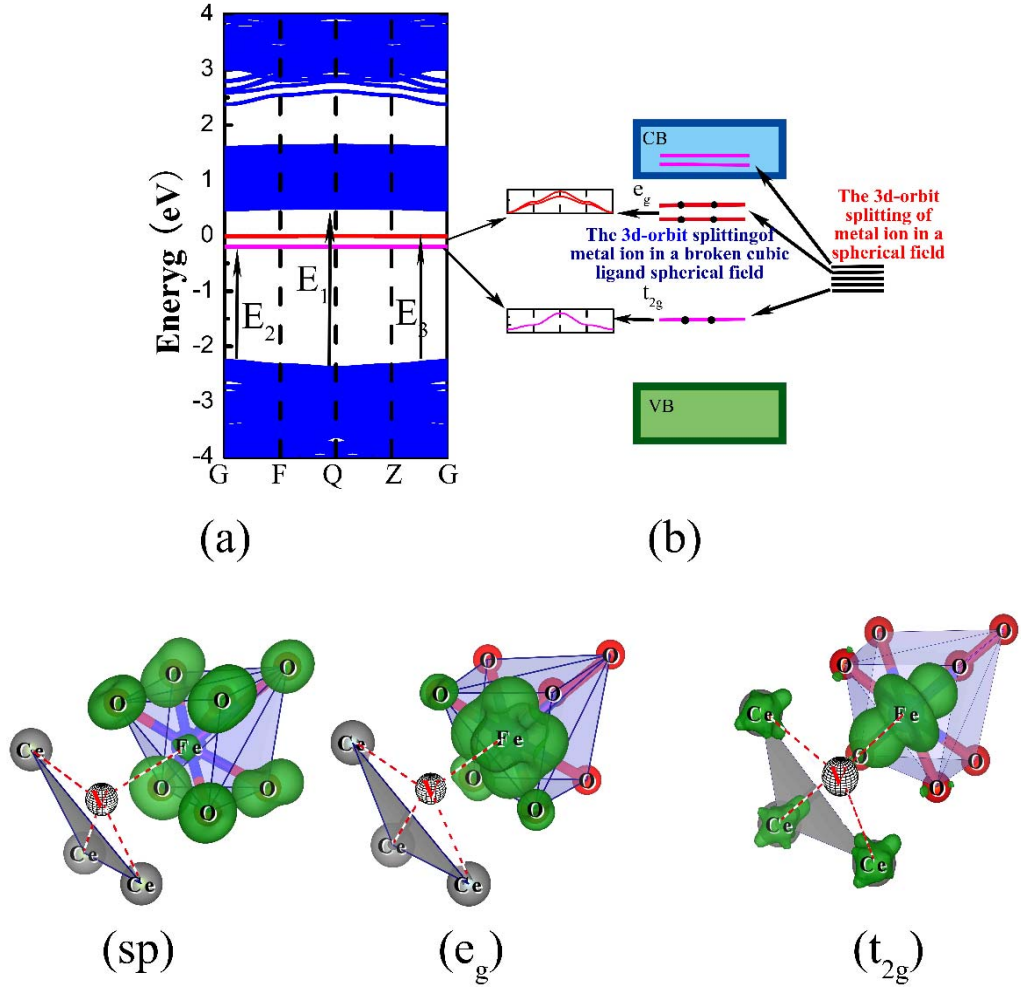

**Fig. 5** (a) Band structure of $Fe_{Ce}$-$V_o$-$CeO_2$. The Fermi level is set at 0 eV. The red and magenta lines denote the $e_g$ and $t_{2g}$ levels, respectively. (b) Schematic plots Fe-3$d$-orbits splitting for $Fe_{Ce}$-$V_o$-$CeO_2$. Bottom panels show the electron and hole density distributions with an isovalue of 0.008 e/Å3: ($sp$), ($e_g$) and ($t_{2g}$) are 4$s$4$p$, $e_g$ and $t_{2g}$ impurity levels of $Fe_{Ce}$-$V_o$-$CeO_2$, respectively.

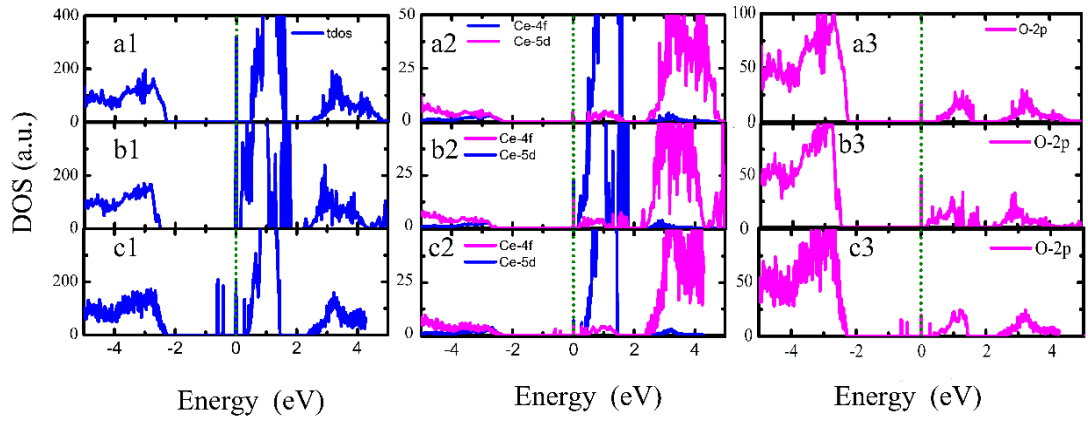

**Fig. 6** DOS and PDOS for Fe-V$_o$-CeO$_2$ (a1-a3), Cr-V$_o$-CeO$_2$ (b1-b3), and Co-V$_o$-CeO$_2$ (c1-c3). The Fermi level is set at 0 eV.

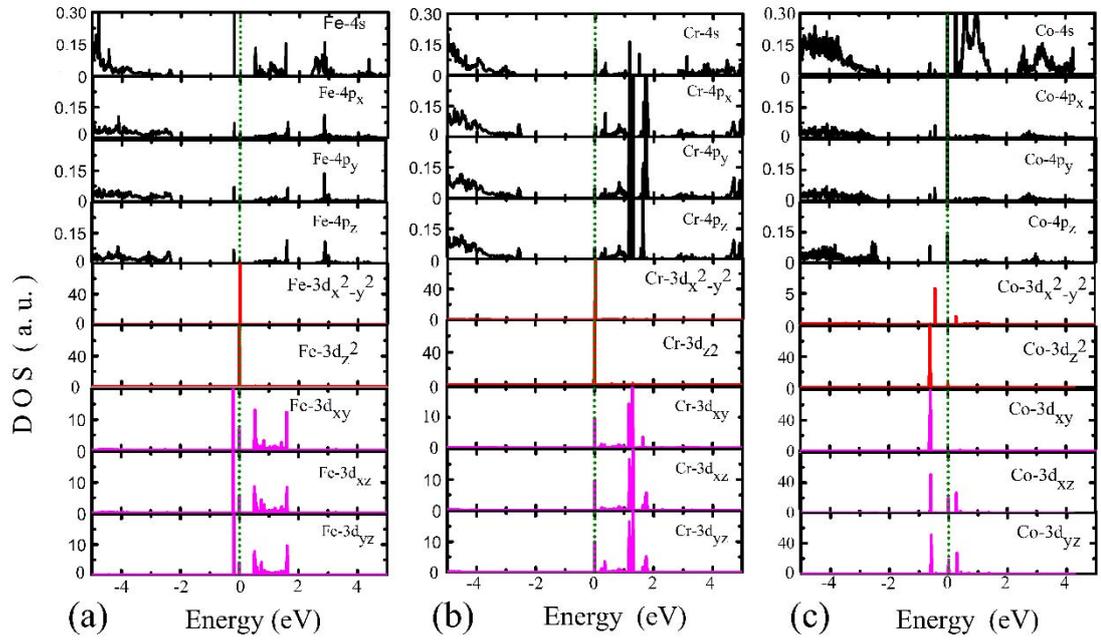

**Fig. 7** PDOS of impurity atoms in $Fe_{Ce}$-$V_o$-$CeO_2$ (a), $Cr_{Ce}$-$V_o$-$CeO_2$ (b), and $Co_{Ce}$-$V_o$-$CeO_2$ (c). The Fermi level is set at 0 eV.

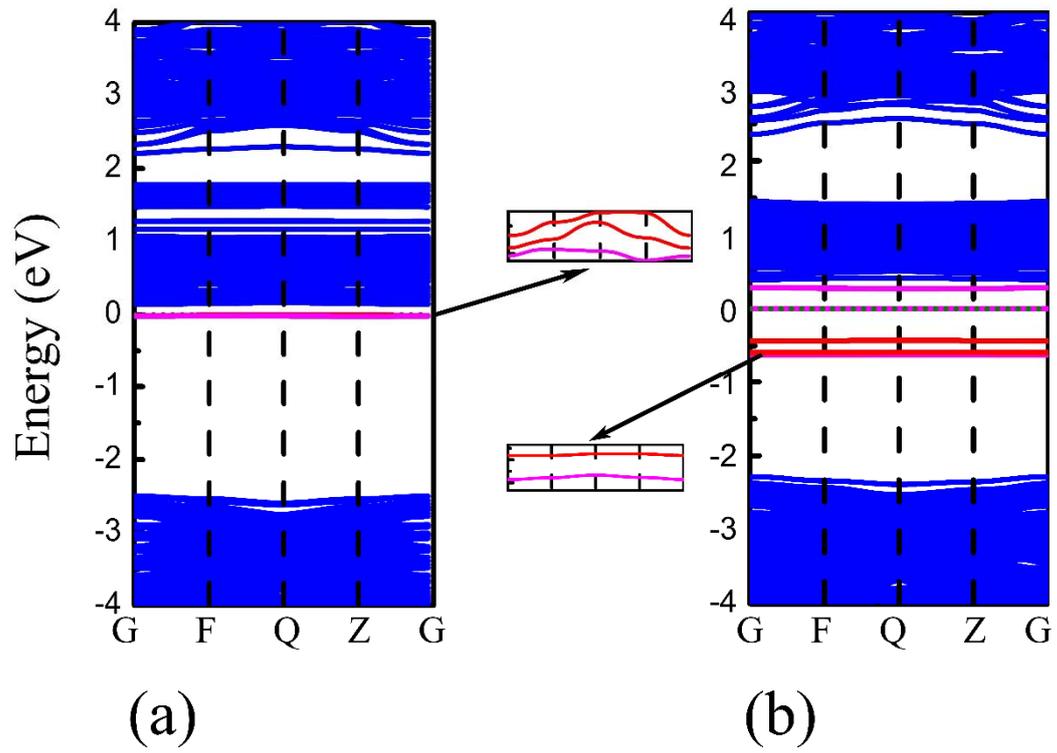

**Fig. 8** Band structures of (a) $Cr_{Ce}$-$V_o$-$CeO_2$ and (b) $Co_{Ce}$-$V_o$-$CeO_2$. The Fermi level is set at 0 eV. The red and magenta lines denote the $e_g$ and $t_{2g}$ impurity levels, respectively.

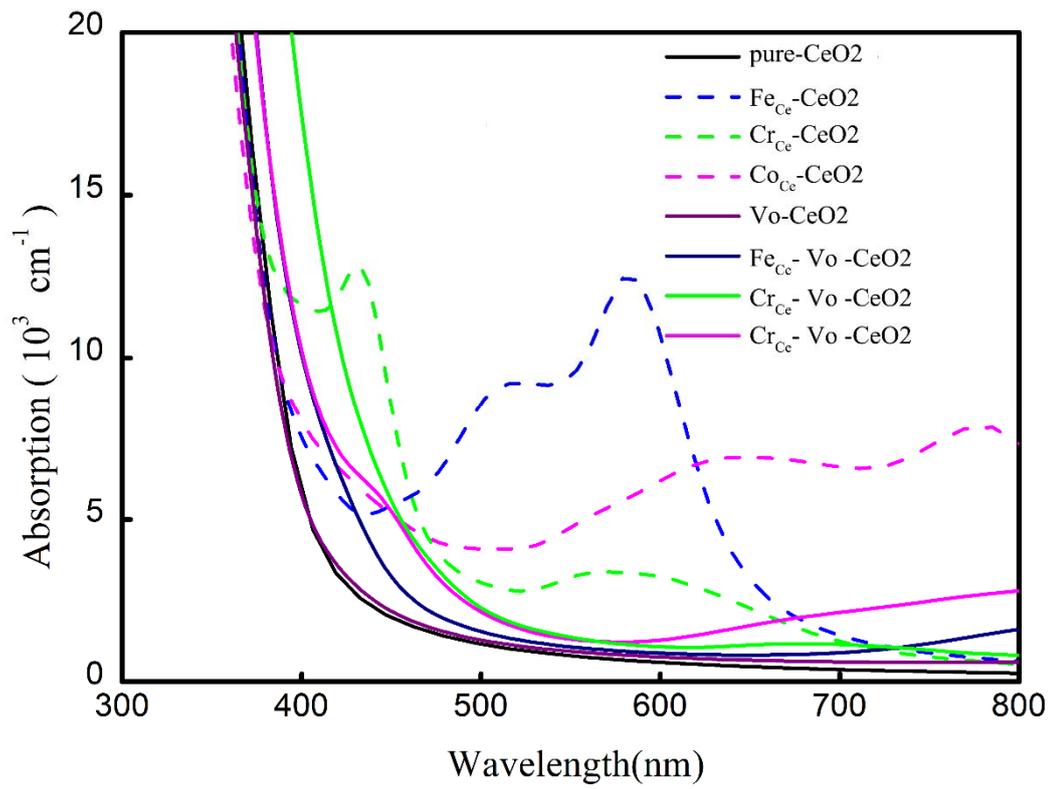

**Fig. 9** Calculated optical absorption spectrum of pure CeO$_2$ and CeO$_2$ doped with transition metal impurity.

# Origin of enhanced visible-light photocatalytic activity of transition metal (Fe, Cr and Co) doped CeO$_2$: Effect of 3$d$-orbital splitting

Ke Yang, Dong-Feng Li, Wei-Qing Huang*, Liang Xu, Gui-Fang Huang#, Shuangchun Wen

*Department of Applied Physics, School of Physics and Electronics, Hunan University, Changsha 410082, China*

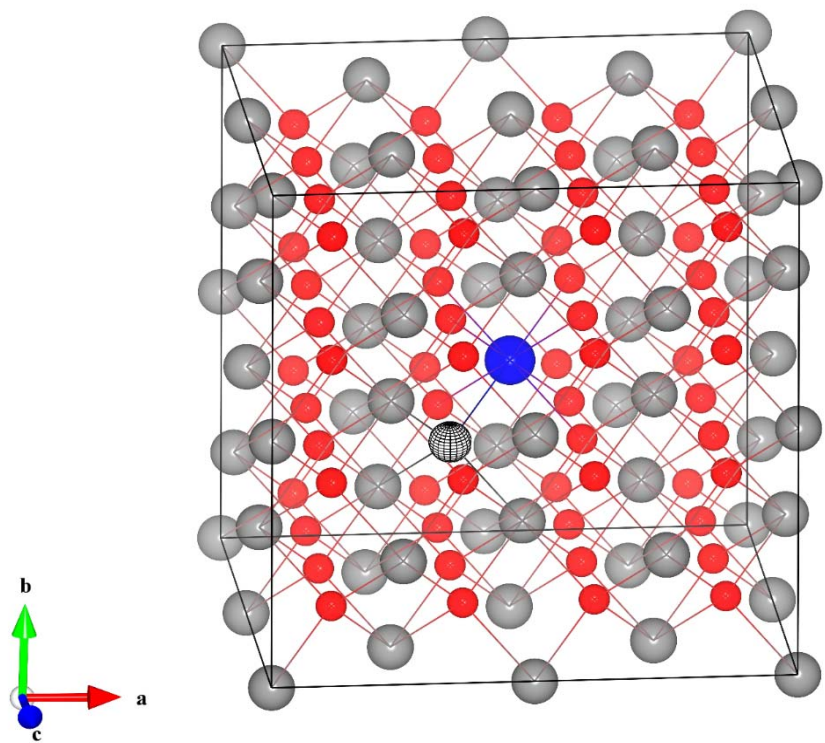

**Fig S1** The geometry structure for 2×2×2 supercell of cubic CeO$_2$. The light gray, red and blue spheres represent Ce, O, and Fe (Cr and Co)atoms, and black hollow sphere denotes an O vacancy.

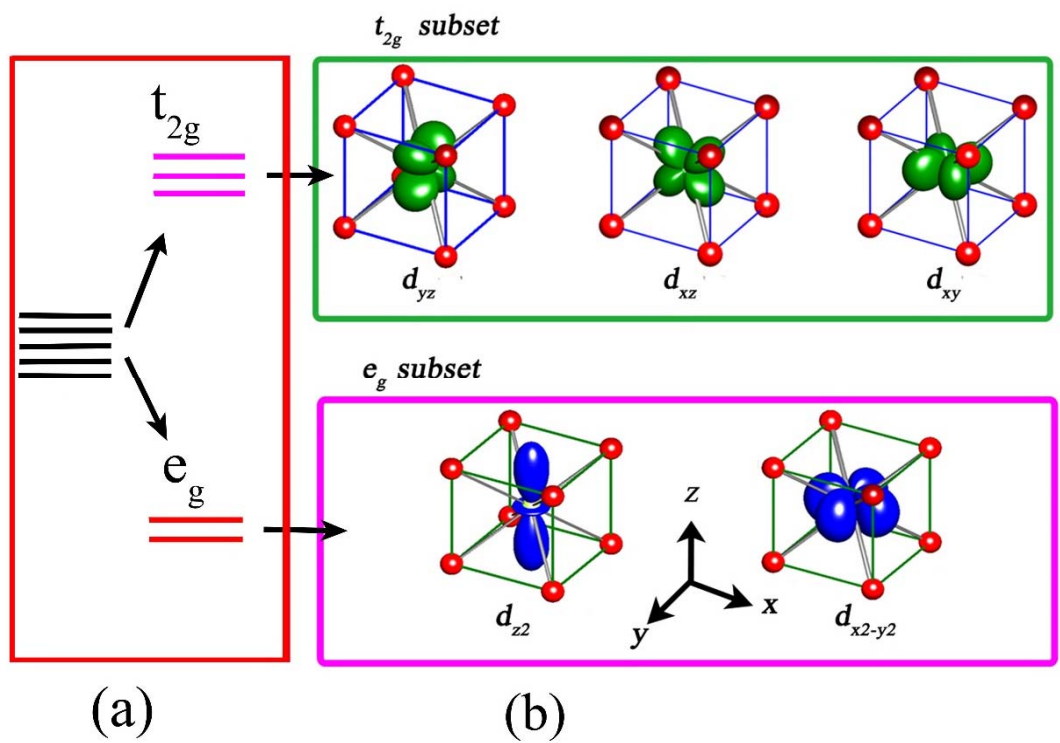

**Fig S2** Schematic plots the 3*d* orbits splitting in ligand field (a) and electron density distribution (b).

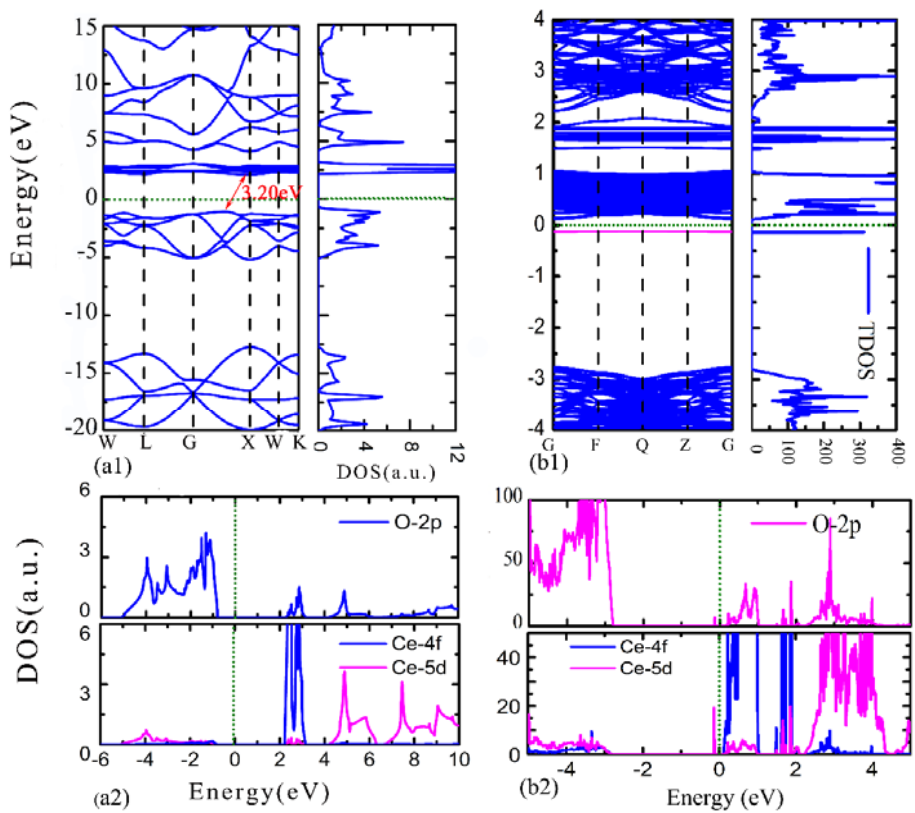

**Fig S3** (a1), (a2) and (b1), (b2) band structure and density of states (DOS, right panel in the upper part) for pure CeO₂ and V_O-CeO₂, respectively; Partial density of states (the lower panel) for pure CeO₂ and V_O-CeO₂, respectively. The dashed lines denote the Fermi level.

The calculated band gap (between the occupied 2$p$ band of O and unoccupied empty $f$ band of Ce) of pure $CeO_2$ is 3.2 eV, which is agreement with the well-established experimental value. The upper part of valence band is dominated by O 2$p$ states, while the narrow band situated just above the Fermi level is mainly due to $f$ states of Ce. The width of the O 2$p$ band is about 4.0 eV, in good agreement with experiment as well as previous calculations. The energy gap between the occupied 2$p$ band of O and unoccupied band of 5$d$ states of Ce, situated above the empty $f$ band of Ce, is about 5.1 eV, which also agrees well with previous calculations.

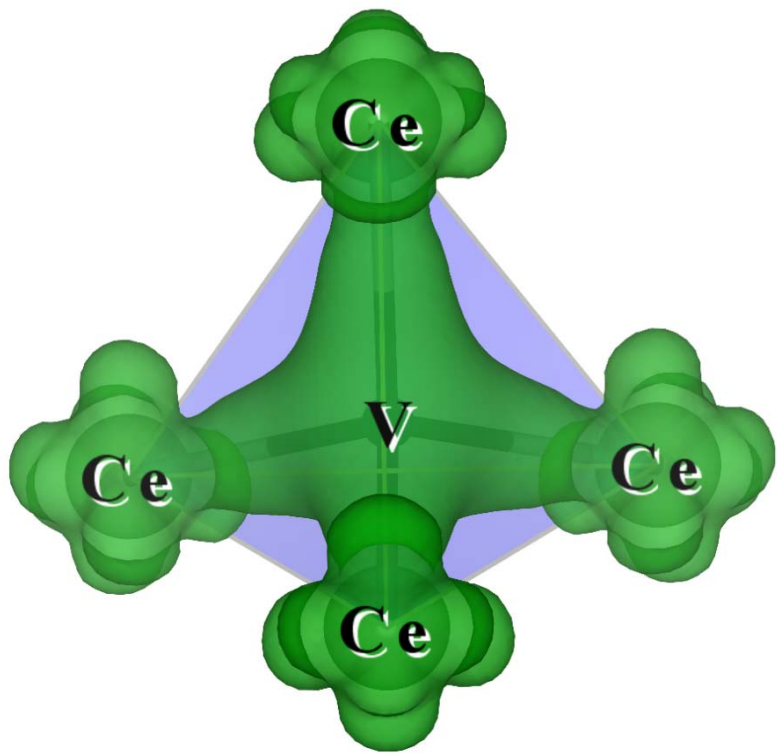

**Fig S4** Plots of the electron and hole density distributions of impurity levels near CB for $V_O$-$CeO_2$, and the isovalue is 0.005e/Å3. Here, V denotes O vacancy.